\documentclass[aps,prl,twocolumn,amsfonts,amssymb,amsmath,floats,floatfix,showpacs,preprintnumbers,superscriptaddress]{revtex4-1}
\usepackage{graphicx}
\usepackage{dcolumn}
\usepackage{bm}
\graphicspath{{C:/Riccardo/Papers/Na2IrO3_K_LDA/Article_images/PRL_images/}} 

\begin{document}

\title{Na$_2$IrO$_3$ as a Novel Relativistic Mott Insulator with a 340\,meV Gap}

\author{R. Comin}
\affiliation{Department of Physics {\rm {\&}} Astronomy, University of British Columbia, Vancouver, British Columbia V6T 1Z1, Canada}
\author{G. Levy}
\affiliation{Department of Physics {\rm {\&}} Astronomy, University of British Columbia, Vancouver, British Columbia V6T 1Z1, Canada}
\affiliation{Quantum Matter Institute, University of British Columbia, Vancouver, British Columbia V6T 1Z4, Canada}
\author{B. Ludbrook}
\author{Z.-H. Zhu}
\author{C.N. Veenstra}
\author{J.A. Rosen}
\affiliation{Department of Physics {\rm {\&}} Astronomy, University of British Columbia, Vancouver, British Columbia V6T 1Z1, Canada}
\author{Yogesh Singh}
\affiliation{Indian\,Institute\,of\,Science\,Education\,and\,Research\,(IISER)\,Mohali, Knowledge\,City, Sector\,81, Mohali\,140306, India}
\author{P.\,Gegenwart}
\affiliation{I. Physikalisches Institut, Georg-August-Universit\"{a}t G\"{o}ttingen, D-37077 G\"{o}ttingen, Germany}
\author{D. Stricker}
\author{J.N. Hancock}
\author{D. van der Marel}
\affiliation{D\'{e}partment\,de\,Physique\,de\,la\,Mati\`{e}re\,Condens\'{e}e, Universit\'{e} de Gen\`{e}ve, CH-1211 Gen\`{e}ve 4, Switzerland}
\author{I.S. Elfimov}
\author{A. Damascelli}
\email{damascelli@physics.ubc.ca}
\affiliation{Department of Physics {\rm {\&}} Astronomy, University of British Columbia, Vancouver, British Columbia V6T 1Z1, Canada}
\affiliation{Quantum Matter Institute, University of British Columbia, Vancouver, British Columbia V6T 1Z4, Canada}

\date{August 24, 2012}

\begin{abstract}
We study Na$_{2}$IrO$_{3}$ by ARPES, optics, and band structure calculations in the local-density approximation (LDA). The weak dispersion of the Ir 5$d$-$t_{2g}$  manifold highlights the importance of structural distortions and spin-orbit coupling (SO) in driving the system closer to a Mott transition. We detect an insulating gap $\Delta_{gap}\!\simeq\!340$\,meV which, at variance with a Slater-type description, is already open at 300\,K and does not show significant temperature dependence even across $T_N\!\simeq\!15$\,K. An LDA analysis with the inclusion of SO and Coulomb repulsion U reveals that, while the prodromes of an underlying insulating state are already found in LDA+SO, the correct gap magnitude can only be reproduced by LDA+SO+U, with $U\!=\!3$\,eV. This establishes Na$_2$IrO$_3$ as a novel type of Mott-like correlated insulator in which Coulomb and relativistic effects have to be treated on an equal footing.
\end{abstract}


\pacs{\vspace{-0.05cm}71.20.Be, 74.25.Jb, 74.25.Gz, 71.15.Mb}

\maketitle

The proposal of an effective $J_{eff}\!=\!1/2$ Mott-Hubbard state in Sr$_{2}$IrO$_{4}$ \cite{Kim2008} came as a surprise since this case departs from the established phenomenology of Mott-insulating behavior in the canonical early 3$d$ transition-metal oxides.\,There,\,the localized nature of the 3$d$ valence electrons is responsible for the small bandwidth $W$, large Coulomb repulsion $U$, and suppression of charge fluctuations \cite{Mott1949,ZSA}. In particular, Sr$_{2}$IrO$_{4}$ appears to violate the $U\!>\!W$ Mott criterion, which for the very delocalized 5$d$ Ir electrons is not fulfilled. It was proposed that the strong spin-orbit (SO) interaction in 5$d$ systems ($\zeta_{SO}\!\simeq\!485$\,meV for Ir \cite{SO}) might lead to instability against weak electron-electron correlation effects, and to the subsequent emergence of a many-body insulating ground state \cite{Kim2008}. However,\,the strong-SO\,limit $J_{eff}\!=\!1/2$ ground-state scenario has recently been put into question \cite{Haskel2012}, and theoretical \cite{Arita2012} and time-resolved optical studies \cite{Hsieh} suggest that the insulating state of Sr$_{2}$IrO$_{4}$ might be closer to a Slater than a Mott type: a band-like insulating state induced by the onset of antiferromagnetic (AF) ordering and consequent band folding at $T_N\!\simeq\!240$\,K (Slater), as opposed to be driven by electron-correlations with an insulating gap already open at temperatures well above $T_N$ (Mott).  

Despite intense experimental and theoretical effort, the nature of the insulating state in the 5$d$ iridates remains highly controversial. This is reminiscent of the situation in 3$d$ oxides, for which the Mott versus band-insulator debate has lasted over four decades \cite{Mott1949,Terakura1984,Fujimori1984,Fujimori1984_bis}. For instance, in the case of the  prototypical AF insulator NiO, this debate was conclusively resolved only after the correlated nature of the insulating state was established based on: ({\it i\,}) the magnitude of the gap as measured by direct and inverse photoelectron spectroscopy (PES/IPES) \cite{Sawatzky_Allen}, much larger than expected from density functional theory (DFT) \cite{Terakura1984}; ({\it ii\,}) its persistence well above the N\'{e}el temperature $T_N$ \cite{Tjernberg1996}; and  ({\it iii\,}) the detailed comparison between DMFT results \cite{Kunes2007} and  momentum-resolved electronic structure as measured by angle-resolved photoelectron spectroscopy (ARPES) \cite{Tjernberg1996,Shen1990}.

To address the nature of the insulating state in iridates, including the role of many-body electron correlations for their extended 5$d$ orbitals and the delicate interplay between $W$, $U$ and SO energy scales, a particularly interesting system is the newly discovered AF insulator Na$_{2}$IrO$_{3}$ \cite{Singh2010}. Starting from a $J_{eff}\!=\!1/2$ model in analogy with Sr$_{2}$IrO$_{4}$, this system was predicted to exhibit quantum spin Hall behavior, and was considered a potential candidate for a topologically insulating state \cite{Shitade2009}. Further theoretical \cite{Jin2009,Bhattacharjee} and experimental \cite{Feng2012} work emphasized the relevance of structural distortions, which lower the local symmetry at the Ir site from octahedral ($O_{h}$) to trigonal ($D_{3h}$). Together with the structure comprised of edge-sharing IrO$_6$ octahedra, this leads to an effective bandwidth  for the Ir 5$d$-$t_{2g}$ manifold of $\sim\!1$\,eV. This potentially puts Na$_2$IrO$_3$ closer than other iridates to the $U\!\sim\!W$ Mott criterion borderline -- and thus to a Mott insulating phase \cite{Jin2009}. Most importantly, its lower $T_N\!\simeq\!15$\,K provides the opportunity of studying the electronic structure well above the long-range AF ordering temperature and -- with the aid of novel DFT calculations -- establishing the nature of its insulating behavior.

In this Letter we present a study of the low-energy electronic structure of Na$_{2}$IrO$_{3}$ by ARPES, angle-integrated PES with in-situ potassium doping, optics, and DFT calculations in the local-density approximation (LDA). The narrow bandwidth of the Ir-5$d$  manifold observed in ARPES highlights the importance of SO coupling and structural distortions in driving the system towards a Mott transition. In addition, at variance with a Slater-type description, the gap is already open at 300\,K and does not show significant temperature dependence even across $T_N\!\simeq\!15$\,K. From the potassium-induced chemical potential shift and complementary optical conductivity measurements, we estimate the insulating gap to be $\Delta_{gap}\!\simeq\!340$\,meV. 
\begin{figure}[t!]
\includegraphics[width=1\linewidth]{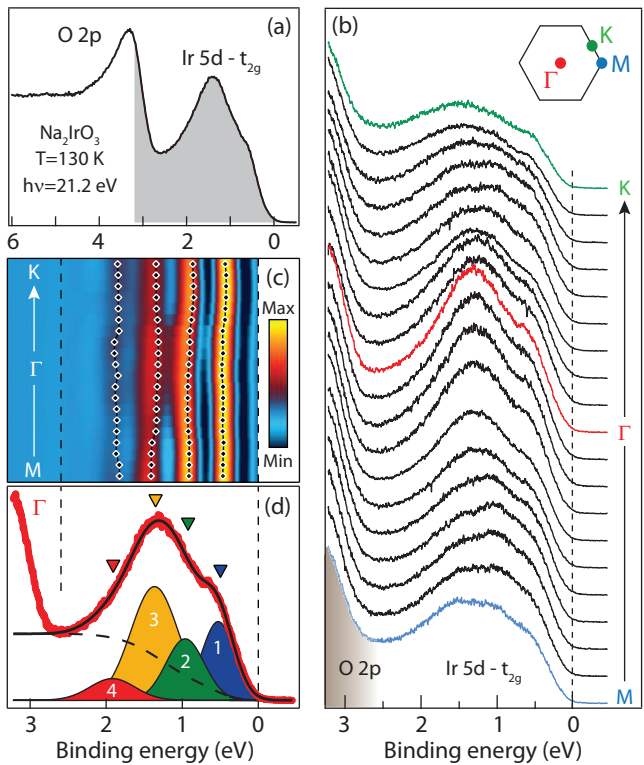}
\caption{\label{fig:ARPES_stack} (color online).  (a) Angle-integrated O and Ir valence-band photoemission spectrum of Na$_2$IrO$_3$; the grey portion is shown in detail in (b-d). (b) ARPES EDCs for the Ir 5d-$t_{2g}$ bands from along $M\!-\!\Gamma\!-\!K$. (c) Negative second derivative of ARPES map highlighting the experimental dispersion; superimposed (black diamonds) are the fit analysis results from (d). (d) Model fit of the $\Gamma$-point EDC with 4 Gaussian peaks for the Ir valence bands (VBs), and a Shirley background \cite{Damascelli_PS}: 4 peaks are necessary to fit the dataset over the full momentum range [matching the number of DOS features from DFT in Fig.\,\ref{fig:LDA_results}(e,f)], with Gaussian lineshapes yielding better agreement than Lorentzians. The average over  $M\!-\!\Gamma\!-\!K$ gives ($E_{VB}$,$\Gamma_{VB}$,$\Delta E_{VB}$) in eV, for peak 1 to 4: (0.50,0.30,0.06); (0.94,0.39,0.08); (1.39,0.44,0.09); (1.89,0.43,0.07).}
\end{figure}
While LDA+SO already returns a depletion of density of states (DOS) at ${E}_{F}$, this only corresponds to a `zero-gap'. The observed $340$\,meV gap value can be reproduced only in LDA+SO+U calculations, i.e. with {\it the inclusion of both SO and U} (with $U\!\simeq\!3$\,eV), establishing Na$_2$IrO$_3$ as a novel {\it relativistic Mott insulator}.

The 130\,K  angle-integrated PES spectrum  \cite{method_ARPES} in Fig.\,\ref{fig:ARPES_stack}(a) shows two broad spectral features belonging to the Ir 5$d$-$t_{2g}$ bands (0-3\,eV binding energy), and to the O 2$p$ manifold (beyond 3\,eV). The insulating character is evidenced by the lack of spectral weight at the chemical potential, which appears to be pinned to the top of the valence band (no temperature dependence is observed in the 130-250\,K range \cite{method_ARPES,supplementary}). Energy distribution curves (EDCs)  measured by ARPES  \cite{method_ARPES} along $M\!-\!\Gamma\!-\!K$ for the Ir 5$d$-$t_{2g}$ bands are shown in Fig.\,\ref{fig:ARPES_stack}(b). The detected features are only weakly dispersing in energy, with the most obvious momentum-dependence being limited to their relative intensity. The electronic dispersion can be estimated from the negative second derivative map in Fig.\,\ref{fig:ARPES_stack}(c), calculated as $ - {\partial}^{2} I ( \mathbf{k}, E ) / {\partial E}^{2} $, and more quantitatively from the fit of EDCs as in Fig.\,\ref{fig:ARPES_stack}(d) (see caption for details). The direct comparison of fit (black diamonds) and second derivative results in Fig.\,\ref{fig:ARPES_stack}(c) yields a good overall agreement in the dispersion of the 4 features (small deviations stem from the peaks' relative intensity variation, which is differently captured by the two methods). We find that the Ir 5$d$-$t_{2g}$ valence band (VB) dispersions do not exceed $\Delta E_{VB}\!\sim\!100$\,meV in bandwidth -- at variance with the generally expected larger hopping amplitude for 5$d$-$t_{2g}$ states. Another remarkable aspect of the Ir $t_{2g}$ bands is their linewidth, with values $\Gamma_{VB}\!=\!\sqrt{2}\sigma_{VB}\!\sim\!300\!-\!450$\,meV. A possible origin might be many-body electron correlation effects as discussed for NiO \cite{Kunes2007}, and strong electron-phonon coupling leading to polaronic behavior in the spectral function \cite{kyle}.

The results in Fig.\,\ref{fig:ARPES_stack} already provide one very important clue: the gap is open well above $T_N\!\simeq\! 15$\,K, which directly excludes a Slater-type, magnetic-order-driven nature for the insulating state. As for the size of the gap, this cannot be readily identified by ARPES since photoemission can locate the valence band, as the first electron-removal state, but not the conduction band which belongs to the electron-addition part of the spectral function \cite{Damascelli_PS}. Alternatively, one can  measure the gap in an optical experiment; however, one needs to discriminate between in-gap states of bosonic character (e.g., phonons, magnons, excitons) and those particle-hole excitations which instead determine the real charge gap. This complication can often hinder the precise identification of the gap edge \cite{Basov2011}. Such complexity underlies the past controversy on NiO: while the 0.3\,eV gap obtained by DFT \citep{Terakura1984} was deemed consistent with optical experiments \citep{Newman1959}, the combination of direct and inverse PES revealed the actual gap value to be 4.3\,eV \citep{Sawatzky_Allen}. This is well beyond the DFT estimate and establishes NiO as a correlated insulator. 
\begin{figure}[t!]
\includegraphics[width=1\linewidth]{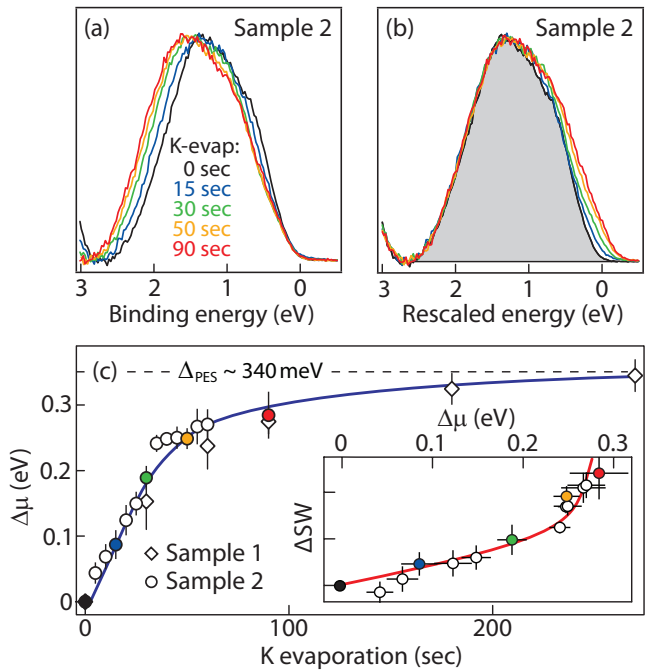}
\caption{\label{fig:Kevap} (color online). (a) Background-subtracted angle-integrated EDCs for selected values of K-exposure [see colored markers in (c)]; the  chemical potential shift $\Delta\mu$ is revealed by the motion of the high binding-energy trailing edge. (b) Same as in (a), but shifted by the corresponding $\Delta\mu$. (c) $\Delta\mu$ vs. K-deposition time for two different samples; in the inset, K-induced low-energy spectral weight $\Delta SW$ as a function of $\Delta\mu$, for Sample 2 only. In panel (c) data and error bars are estimated from the comprehensive analysis of both O and Ir trailing edges and peak positions \cite{supplementary}; also note that blue and red curves in (c) and its inset are both a guide-to-the-eye.}
\end{figure}
Here, for the most conclusive determination of the insulating gap magnitude, we use angle-integrated PES with in-situ doping by potassium deposition, as well as optics. A quantitative agreement between the two probes would provide validation against possible artefacts \cite{artefacts}.

To estimate the energy of the first electron-addition states and the DOS gap between valence and conduction bands, we start by doping carriers (i.e., electrons) across the gap by in situ potassium deposition on the cleaved surfaces, and then follow the shift in chemical potential 
by angle-integrated PES. The results are summarized in Fig.\,\ref{fig:Kevap} for K-evaporation performed at 130\,K on two different freshly cleaved surfaces \cite{supplementary}. The most evident effect is the shift towards higher binding energy of both Ir and O valence bands, as shown in Fig.\,\ref{fig:Kevap}(a) for the Ir 5$d$-$t_{2g}$ manifold \cite{supplementary}. 
\begin{figure}[b!]
\includegraphics[width=1\linewidth]{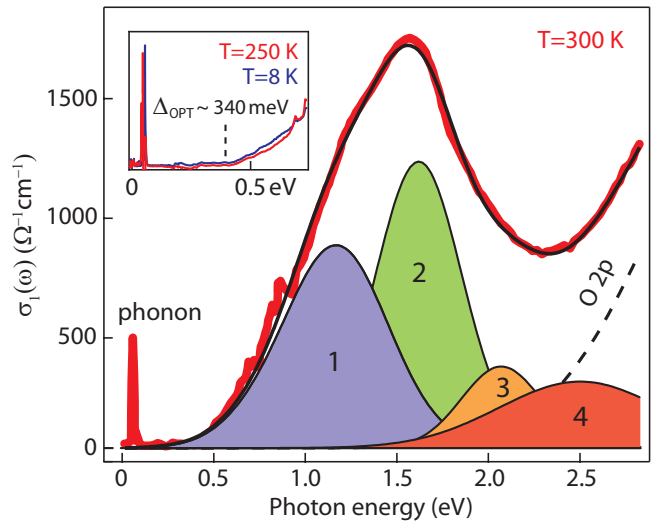}
\caption{\label{fig:OC} (Color online). Optical conductivity data (red line), together with the simulated Ir 5$d$-$t_{2g}$ joint particle-hole DOS (black line), and its individual components from the simultaneous fit of ARPES and optical data [colors and labels are consistent with the valence band features in Fig.\,\ref{fig:ARPES_stack}(c), which represent the initial states of the lowest-energy optical transitions]. Inset: temperature dependence of the gap edge.}
\end{figure}
\begin{figure*}[t!]
\includegraphics[width=1\linewidth]{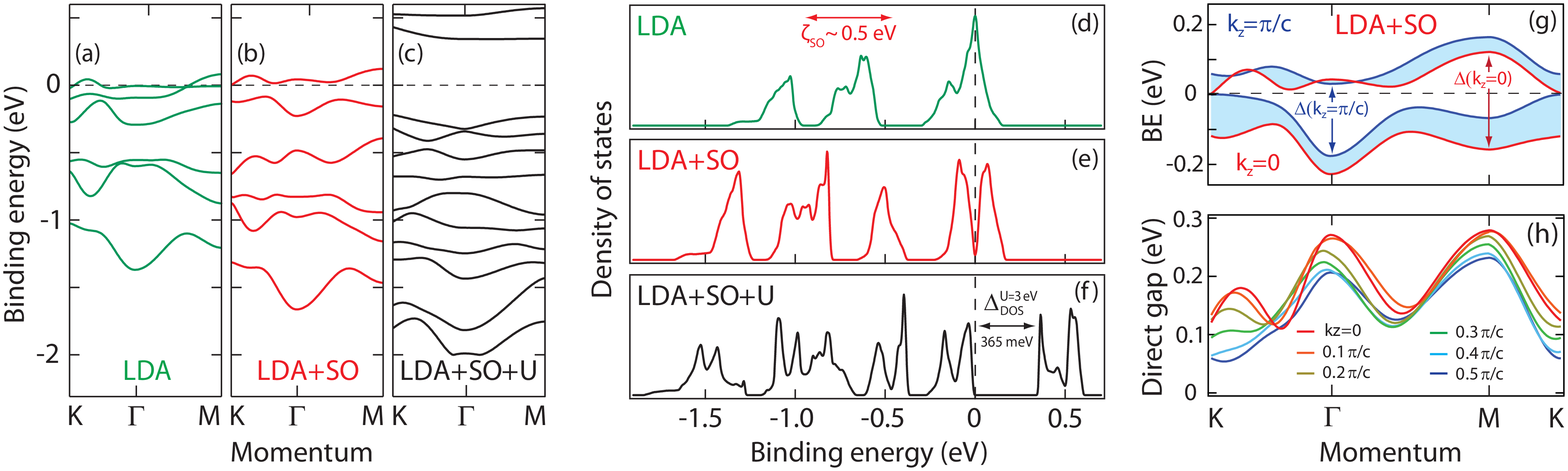}
\caption{\label{fig:LDA_results} (color online). (a-c) Ir 5d-$t_{2g}$ band structure ($k_z\!=\!0$), and (d-f) corresponding DOS, obtained with LDA, LDA+SO, and LDA+SO+U ($U\!=\!3$\,eV, $J_H\!=\!0.6$\,eV \cite{Ueff}). (g) $k_z$ dispersion, for the last occupied and first unoccupied Ir 5d-$t_{2g}$ bands from LDA+SO, as indicated by the filled region between the $k_z\!=\!0$ and $\pi/c$ extreme lines. While the filled areas overlap, resulting in a vanishing indirect gap (i.e. $\Delta k\!\neq\!0$), the direct gap (i.e. $\Delta k\!=\!0$) between valence and conduction bands is finite for all $k_z$ and $k_{\parallel}$. (h) LDA+SO direct gap distribution along $M-\Gamma-K$, for different $k_z$ in the 3-dimensional Brillouin zone.}
\end{figure*}
This arises from the (equal and opposite) shift of the chemical potential $\Delta\mu $ when electrons donated by potassium are doped into the system; after an initial rapid increase, $ \Delta \mu $ saturates at $\sim\!340$\,meV [Fig.\,\ref{fig:Kevap}(c)]. When the K-deposited spectra are shifted in energy by the corresponding $\Delta\mu $ so that their high binding-energy  trailing edges match the one of the fresh surface [Fig.\,\ref{fig:Kevap}(b)], one can observe the emergence of additional spectral weight ($SW$) in the region close to and above $E_F$. This low-energy K-induced spectral weight, $\Delta SW$, can be computed as: 
\begin{equation}
\Delta SW =\! \int d k \int_{-1eV}^{{E}_{F}^+} d E\, [I (k, E, {x}_{{K}^{+}}) - I (k, E, 0)] \, ,
\label{eq:SW}
\end{equation}
\noindent 
where $I(k,\omega)$ is the PES intensity, $ {x}_{{K}^{+}} $ represents the K-induced surface doping, and $E_F^+$ is the Fermi energy of the K-doped surface, which moves progressively beyond the undoped-surface $E_F$. The evolution of $\Delta SW$ plotted versus $\Delta\mu$ in the inset of Fig.\,\ref{fig:Kevap}(c) evidences an approximately linear SW increase up to $\Delta\mu\!\simeq\!200-250$\,meV, followed by a steeper rise once the saturation value $\Delta\mu\!\simeq\!340$\,meV is being approached. This behavior can be understood as due to the initial filling in of in-gap defect states -- either pre-existing or induced by K deposition -- which makes the jump of the chemical potential not as sudden as for a clean insulating DOS. Only when electronic states belonging to the Ir 5$d$-$t_{2g}$ conduction band are reached one observes the saturation of $\Delta\mu$ and the more pronounced increase in  $\Delta SW$. This combined evolution of chemical potential shift and spectral weight increase points to a DOS insulating gap $\Delta_{PES}\!\simeq\!340$\,meV.

Turning now to the optical conductivity  data \cite{method_optics}, in Fig.\,\ref{fig:OC} we observe an insulating behavior with an absorption edge starting at 300-400\,meV at 300\,K, with negligible temperature dependence down to 8\,K and thus also across $T_N\!\simeq\!15$\,K (see inset). We can fit the results using a joint DOS with Gaussian peaks for the conduction band (CB) and each of the 4 VBs (see caption of Fig.\,\ref{fig:ARPES_stack}): 
\begin{equation}
J (E) \propto \sum_{i=1}^4 \int dE' {A}_{i} \, {G}_{C\!B} (E' + E) \, {G}_{V\!Bi} (E') \,.
\label{eq:JDOS}
\end{equation}

\noindent
Here the prefactors $ {A}_{i} $ represent the optical transition strengths, with the band index {\it i} running over the 4 VB features extracted from the ARPES data in Fig.\,\ref{fig:ARPES_stack}, and are left free.  $ J(E)$ provides an excellent fit to the optical data in Fig.\,\ref{fig:OC}, and a least-squares analysis returns ${E}_{CB}\!\simeq\!680$\,meV for the location of the conduction band above $E_F$, with a width $\Gamma_{CB}\!=\!\sqrt{2}\sigma_{CB}\!\simeq\!160$\,meV (a value in agreement with DFT, as shown later). The consistency of the combined PES-optical conductivity analysis is confirmed by the optical gap obtained from the onset of the simulated conduction band, $\Delta_{OPT}\!=\!E_{CB}^{onset}$. Following Ref.\,\onlinecite{kyle}, the latter is estimated as $E_{CB}^{onset}\!=\!{E}_{CB}\!-\!3\sigma_{CB}$, leading to $\Delta_{OPT}\!\simeq\!340$\,meV. This matches $\Delta_{PES}$ from the K-induced $\Delta \mu$ saturation in PES, providing a definitive estimate for the insulating gap, $\Delta_{gap}\!\simeq\!340$\,meV.  

With a gap much smaller than in typical Mott insulators, discriminating between correlated and band-like insulating behavior in Na$_2$IrO$_3$ requires a detailed comparative DFT analysis \citep{LDA}. Unlike the case of Sr$_2$IrO$_4$, the $t_{2g}$ degeneracy and bandwidth in Na$_2$IrO$_3$ are affected by structural distortions and the presence of Na in the Ir plane. This is revealed by calculations we have performed for a distortion- and Na-free hypothetical IrO$_2$ parent compound: while in IrO$_2$ the individual $t_{2g}$-dispersions are as wide as $\sim\!1.6$\,eV, the Na-induced band folding in Na$_2$IrO$_3$ opens large band gaps, leading to much narrower $t_{2g}$-subbands ($\sim\!100$\,meV). Even though this accounts well for the narrow bandwidth observed in ARPES, the material is still metallic in LDA with a high DOS at the Fermi level [Fig.\,\ref{fig:LDA_results}(a,d)], at variance with the experimental findings. Remarkably, when SO is switched on in LDA+SO a clear gap opens up at $E_F$ in the $k_z\!=\!0$ band dispersion  [Fig.\,\ref{fig:LDA_results}(b)], although not in the DOS where only a {\it zero-gap} can be observed [Fig.\,\ref{fig:LDA_results}(e)]. A closer inspection of the LDA+SO dispersion for first occupied and unoccupied bands -- versus both $k_{\parallel}$ and $k_z$ in Fig.\,\ref{fig:LDA_results}(g) -- reveals that the lack of a DOS gap stems from the overlap of VB and CB at $\Gamma$ and $K$ points for different $k_z$ values. In other words, while the {\it direct} gap ($\Delta k\!=\!0$) is non-zero and ranges from a minimum of 54\,meV to a maximum of 220\,meV over the full Brillouin zone [Fig.\,\ref{fig:LDA_results}(h)], the {\it indirect} DOS gap ($\Delta k\!\neq\!0$) is vanishing. This is still in contrast with the experimentally determined band-gap magnitude $\Delta_{gap}\!\simeq\!340$\,meV \cite{gap}. 

The disconnection between insulating behavior and onset of AF ordering, together with the quantitative disagreement between observed and calculated gap even in LDA+SO, reveal that Na$_2$IrO$_3$ cannot be regarded as either a Slater or a band insulator. Also, given the narrow $t_{2g}$ bandwidths ($\sim\!100$\,meV), one might expect the system to be even more unstable against local correlations than anticipated. Indeed, a good overall agreement with the data is found in LDA+SO+U, for $U\!=\!3$\,eV and $J_H\!=\!0.6$\,eV \cite{Ueff}: this returns a gap value $\Delta_{DOS}^{U\!=\!3eV}\!\simeq\!365$\,meV [Fig.\,\ref{fig:LDA_results}(f)] close to the experimental $\Delta_{gap}$, and a 2\,eV energy range for the Ir 5d-$t_{2g}$ manifold [Fig.\,\ref{fig:LDA_results}(c)] matching the spectral weight distribution in Fig.\,\ref{fig:ARPES_stack} (note that a doubling of bands is seen in LDA+SO+U due to the imposed AF ordering, but is of no relevance here \cite{LDA}). At a first glance, $U\!=\!3$\,eV might seem a large value for 5$d$ orbitals; however, in the solid, the effective reduction of the atomic value of U strongly depends on the polarizability of the surrounding medium,  which is the result of many factors, \textit{in primis} the anion-cation bond length \cite{Meinders1995}. In this perspective, the value we found is not unreasonable, and is also consistent with the existence of local moments above $T_N$ revealed by the Curie-Weiss magnetic susceptibility behavior with $\theta\!\simeq\!-120$\,K \cite{Singh2010}. 

Our findings point to a Mott-like insulating state driven by the delicate interplay between $W$, $U$, and SO energy scales, in which  co-participating structural distortions also play a crucial role. This establishes Na$_2$IrO$_3$, and possibly other members of the iridate family, as a novel type of correlated insulator in which Coulomb (many-body) and relativistic (spin-orbit) effects cannot be decoupled, but must be treated on an equal footing.

We acknowledge S. Bhattacharjee and G.A. Sawatzky for discussions, D. Wong and P. Dosanjh, for technical assistance. This work was supported by the Max Planck -- UBC Centre for Quantum Materials, the Killam, Sloan, CRC, and NSERC's Steacie Fellowship Programs (A.D.), NSERC, CFI, and CIFAR Quantum Materials.


\begin{thebibliography}{38}
\bibitem [{met()}]{method_optics}%
  \BibitemOpen
  \href@noop {} {\bibfield  {journal} {The complex optical
  conductivity was obtained in the 8-300\,K temperature range using combined
  reflectivity and ellipsometry measurements on the (001) surface of a
  freshly-cleaved crystalline platelet}\ }\BibitemShut
  {NoStop}%
\end{thebibliography}

\providecommand{\noopsort}[1]{}\providecommand{\singleletter}[1]{#1}%

\end{document}